\documentclass[draftcls,onecolumn,a4paper]{IEEEtran}

\IEEEoverridecommandlockouts

\usepackage{todonotes}
\usepackage{amsmath}
\usepackage{amsthm}
\usepackage{amsfonts}
\usepackage{amssymb}
\usepackage{graphicx}
\usepackage{color}
\usepackage{booktabs}
\usepackage[usenames,dvipsnames]{pstricks}
\usepackage{epsfig}
\usepackage{rotating,graphics,psfrag}
\usepackage{mathtools}

\newtheorem{mytheorem}{Theorem}
\newtheorem{mycorollary}{Corollary}
\newtheorem{mylemma}{Lemma}

\theoremstyle{definition}
\newtheorem{mydefinition}{Definition}
\newtheorem{myassumption}{Assumption}

\theoremstyle{remark}

\renewcommand{\vec}[1]{\boldsymbol{#1}}

\newcommand{\set}[1]{\mathcal{#1}}

\newcommand{\sphere}{\mathbb{S}}

\newcommand{\Complex}{\mathbb{C}}
\newcommand{\Ex}[1]{\mathbb{E}\left[ {#1} \right]}

\newcommand{\Reals}{\mathbb{R}}

\newcommand{\multi}[2]{\genfrac{}{}{0pt}{}{#1}{#2}}
\newcommand{\ve}{\text{vec}}
\newcommand{\SINR}{\text{SINR}}
\newcommand{\Prob}[1]{\textnormal{Pr}\left\{ #1 \right\}}

\newif\ifcomments
\commentsfalse
\usepackage{etex} 
\usepackage[color]{changebar}
\usepackage[normalem]{ulem}

\ifcomments
\presetkeys{todonotes}{inline}{}
\newcommand{\removed}[1]{\cbstart\removedfragile{#1}\cbend{}}
\newcommand{\removedfragile}[1]{{\color{red}{\sout{#1}}}{}}

\newcommand{\com}[1]{\todo[author=Comment,inline,color=yellow!40,caption={}]{#1}}
\newcommand{\todoSla}[1]{\todo[author=Task  S\l{}awomir,inline,color=red!70,caption={}]{#1}}
\newcommand{\todoJan}[1]{\todo[author=Task  Jan,inline,color=green!40,caption={}]{#1}}
\newcommand{\todoPet}[1]{\todo[author=Task  Peter,inline,color=green!40,caption={}]{#1}}
\newcommand{\note}[1]{\todo[author=Note,inline,color=blue!40,caption={}]{#1}}

\else

\newcommand{\removed}[1]{}

\newcommand{\cosl}[1]{}
\newcommand{\note}[1]{}
\newcommand{\todoSla}[1]{}
\newcommand{\todoJan}[1]{}
\newcommand{\todoPet}[1]{}
\newcommand{\com}[1]{}
\fi

\title{Compressive Rate Estimation with Applications to Device-to-Device Communications} 

\author{ 
\IEEEauthorblockN{Jan Schreck, Peter Jung and S\l{}awomir Sta\'nczak}
  \IEEEauthorblockA{\\
    Technische Universit\"at Berlin \\
Communications and Information Theory Group\\
Einsteinufer 25, D-10587 Berlin, Germany\\
    Email: jan.schreck@gmail.com, peter.jung@tu-berlin.de,  slawomir.stanczak@tu-berlin.de}
 \thanks{ 
   Peter Jung is supported by  the {\em   Deutsche
 Forschungsgemeinschaft (DFG) }  under grant  JU 2795/2-1.
 }
}

\IEEEspecialpapernotice{(Preprint)}

\begin{document}
\maketitle

\begin{abstract}
  We develop a framework that we call compressive rate estimation. We
  assume that the composite channel gain matrix (i.e. the matrix of
  all channel gains between all network nodes) is compressible which
  means it can be approximated by a sparse or low rank representation.
  We develop and study a novel sensing and reconstruction protocol for
  the estimation of achievable rates.  We develop a sensing protocol
  that exploits the superposition principle of the wireless channel
  and enables the receiving nodes to obtain non-adaptive random
  measurements of columns of the composite channel matrix.  The random
  measurements are fed back to a central controller that decodes the
  composite channel gain matrix (or parts of it) and estimates
  individual user rates. We analyze the rate loss for a linear and a
  non-linear decoder and find the scaling laws according to the number
  of non-adaptive measurements. In particular, if we consider a system
  with $N$ nodes and assume that each column of the composite channel
  matrix is $k$ sparse, our findings can be summarized as follows. For
  a certain class of non-linear decoders we show that if the number of
  pilot signals $M$ scales like $M \sim k \log(N/k)$, then the rate
  loss compared to perfect channel state information remains
  bounded. For a certain class of linear decoders we show that the
  rate loss compared to perfect channel state information scales like $1/\sqrt{M}$.
\end{abstract}





\section{Introduction}
\label{sec:intro}


Device-to-device (D2D) communication has evolved as one of the key
technology enablers for 5G wireless systems (``Beyond 2020 Networks'')
\cite{METISD1.1}. The basic idea of D2D communication is to establish
direct short-distance communication links between pairs of suitably
selected wireless devices so that there is no need for long-distance
transmissions to and from base stations (BS). Exploiting direct
communication between nearby devices has a huge potential for boosting
the performance of cellular networks \cite{Doppler1} and improving the
service quality of proximity based applications \cite{Corson}. In
addition, D2D communication makes some new exciting location-based
services and applications possible.  

The main potential advantages of D2D communications stem from the
proximity-, reuse-, and hop gains that can be summarized as follows
\cite{Fodor12}:
\begin{itemize}
\item Coverage improves since direct D2D links\footnote{We refer the
    reader to Section \ref{sec:setup} for more details about the
    terminology used throughout the paper.} can be used to fill
  coverage holes;
\item Capacity enhances due to the reuse of radio resources of the
  supporting cellular layer by multiple D2D links \cite{Andrews};
\item Energy efficiency increases since transmit powers can be reduced
  without deteriorating the capacity \cite{Belleschi:11};
\item Achievable peak rates increase and end-to-end latencies decrease
  due to proximity and hop gains.
\end{itemize} 

D2D communication has been extensively studied in the context of
ad-hoc networks, in which wireless devices utilize unlicensed spectrum
resources with no or strictly limited assistance from a fixed network
infrastructure. Such solutions are not suitable for general purpose
wireless applications due to the lack of quality-of-service (QoS)
guarantees to D2D links \cite{SRK2003_cross}. This is also true in the
case of other approaches to D2D communication that are based on the
concept of cognitive radio and dynamic/opportunistic spectrum access
\cite{Kaufman-1301-6980}. Therefore, these approaches have found
limited acceptance in the standardization bodies.

In order to overcome the limits of unassisted ad-hoc networking
technologies and opportunistic spectrum access technologies based on
spectrum sensing, researchers have recently turned their attention
towards network-assisted D2D communication, which promises more
efficient spectrum utilization, QoS support and higher reliability,
while providing D2D discovery support, synchronization and security
\cite{Doppler1,Fodor12,Andrews}. In particular, the design aspects of
D2D communication are currently discussed in 3GPP, where the
feasibility and the architecture enhancements of so called proximity
services (ProSe) are under discussion \cite{3GPP2013},
\cite{3GPP2014}. Thereby, D2D links can operate in in-band mode and
out-band mode. While the in-band D2D mode utilizes the same spectral
resources as cellular users that transmit their data via base stations
in the traditional cellular mode, the out-band D2D mode allocates
cellular users and D2D links to different frequency bands. We focus on
in-band D2D communication and assume that all users are in-coverage,
which means that each user is connected to some base
station.\footnote{Nonetheless, we point out that most of the proposed
  methods and concepts can be extended to enable D2D communication for
  out-of-coverage users.} As an underlay to cellular networks, in-band
D2D communication can be seen as a network-assisted interference
channel, in which D2D transmissions reuse cellular resources while
being assisted by base stations.



Despite key advantages, network-assisted D2D communication also poses
some fundamental challenges including transmission mode selection,
robust interference management and feedback design. The underlying
problems are aggravated by the lack of channel state information (CSI)
at different locations in a network.\footnote{Notice that CSI is used
  in a broad sense here and does not necessarily mean the full channel
  knowledge. In particular, CSI may also refer to the information
  about achievable rates.} There is in particular a vital need for
timely and accurate CSI that can be used by the network controller to
facilitate reliable D2D discovery and QoS-aware scheduling.  In other
words, when establishing D2D links and allocating cellular resources
to them, the network controller should have enough CSI to ensure that
the QoS demands of all cellular and D2D users (e.g. expressed in terms
of some minimum data rate requirements) are guaranteed once in-band
D2D links are established. While being highly valuable, CSI is not for
free and must be obtained as efficient as possible without consuming
to much scarce radio resources. In \cite{schreck2013channel} the
authors used methods from compressed sensing to acquire channel state
information at the central controller of two-hop network. 

\subsection{Our Contribution}
\label{sec:intro_contrib}

This paper contributes towards the development of measurement-based
feedback protocols, with the goal of enabling a network controller to
acquire the required CSI in a highly efficient way.  
Such protocols need to perform the following steps
\cite{Belleschi2013}:
\begin{itemize}
\item \emph{Spectral resource management}: The BS assigns cellular
  users to the available spectral resources. This step is performed in
  any cellular network with centralized resource management, e.g.,
  3GPP LTE.
\item \emph{D2D discovery and mode selection}: The BS detects wireless
  devices that are in proximity to each other (D2D discovery) and
  decides if a device should operate in cellular mode or D2D mode
  (mode selection).
\item \emph{Pairing}: The network controller decides if one or more
  D2D links share a spectral resource with some cellular user.
\end{itemize}

The focus of this paper is on D2D discovery -- also called proximity
discovery -- and on pairing, which is a part of scheduling decisions
that assign resources to cellular users and D2D links.
Both tasks -- D2D discovery and pairing -- are entirely carried out by
a network controller where enough CSI is needed for robust
decisions. Assuming D2D communication as an underlay to a cellular
network, we address the problem of \emph{reliable D2D discovery and
  pairing} based on compressed and quantized channel measurements. We
develop and study a novel
sensing and reconstruction strategy (protocol) for the estimation of
achievable rates, which
we call compressive rate estimation. 
The proposed protocol combines the estimation from
compressed measurements with coded access to reduce the number of
pilot-based measurements that need to be fed back to estimate the
achievable rates and to make timely and robust \emph{QoS-aware
  decisions}.
By using the concept of coded access we are able to  exploit
collisions in an interference channel to obtain compressed
non-adaptive measurements from linear random projections (e.g.  analog
coding of \cite{Goldenbaum:Stanczak:11a} can be used for this
purpose).  To estimate the rates, we apply methods from compressed sensing and
sparse approximation \cite{candes2006stable}.  Since a major drawback
of compressed sensing based techniques is that they require highly
complex decoders, we also consider linear estimation methods which
require significantly less complexity
\cite{davenport2006detection}. As we will see, the advantages of the
proposed protocol are three-fold.  First, by applying the concept of
coded access, we are able to significantly reduce the pilot
contamination in the network. 
Second, the feedback overhead is reduced since significantly fewer
measurements need to be quantized and fed back. Third, most of the
complexity required to estimate the achievable rates is imposed on the
network controller.

\subsection{Notation}

The element in the $i$-th row and $j$-th column of a matrix $\vec X$
is given by $[\vec X]_{i,j} = x_{i,j}$, similarly, the $i$-th element
of a vector $\vec x$ is given by $x_i$. The conjugate transpose of a
matrix $\vec X$ is $\vec X^H$.  For vectors the $\ell_p$--norm is
given by $\|\vec x\|_{\ell_p} = \left(\sum_i \vec x_i^p
\right)^{1/p},p\geq 1$. For matrices the Schatten-$p$ norm is given by
$\|\vec X\|_{s_p} = \left(\sum_i \sigma_i^p(\vec X) \right)^{1/p}$
where $\{ \sigma_i(\vec X) \}_{i}$ are the singular values of the
matrix $\vec X$ in decreasing order. The operator $\vec x=\ve(\vec X)$
stacks the columns of the matrix $\vec X$ in a large column vector
$\vec x$. The support $\text{supp}(\vec x)$ of a vector $\vec x$ is
the index set of its non-zero elements. 
The $N\times N$ identity matrix is denoted as $\vec I_N$ and its
$i$-th column is defined as $\vec e_{i}$.  Tuples are denoted by
calligraphic letters and the $i$-th element of tuple $\set X$ is given
by $\set X_i$.  The real numbers are defined as $\Reals$ and the
complex number are $\Complex$.

\section{System Model}
\label{sec:setup}



We consider a cellular network with a large number of wireless devices
and multiple base stations
that 
are controlled by a (central) network controller. We assume there are
$N>1$ transmitters that wish to establish communication links over the
(wireless) channel to transfer independent data to $N$
receivers.\footnote{For simplicity, the reader may assume
  unidirectional communication links throughout the paper but we point
  out that the results can be straightforwardly extended to
  bidirectional links.} Communication links between the wireless
devices and the base stations are referred to as \emph{cellular
  users}, while the term \emph{D2D user} or, equivalently, \emph{D2D
  link} is used to refer to a communication link between two wireless
devices.
The users as well as the corresponding transmitters and receivers are
indexed in an arbitrary but fixed order with indices taken from the
set $\set N = \{ 1,2,\ldots N\}$.\footnote{We also use $\set N$ to
  refer to transmitters, receivers and transmissions (i.e., users
  scheduled for transmissions). According to this, transmission $i \in
  \set N$ is the transmission from transmitter $i\in\set N$ to
  receiver $i\in\set N$} A subset $\set N_1 \subseteq \set N$ is used
to denote cellular users so that the remaining users with indices in
$\set N\setminus\set N_1$ are \emph{potential} D2D users. The cellular
users are assumed to have been scheduled for (cellular) transmissions
in the downlink channel.


A frequency-division multiple access (FDMA) technique such as OFDMA
(OFDMA: orthogonal frequency-division multiple access) together with a
time-division multiple access (TDMA) technique is used to divide the
available bandwidth and time in a number of mutually orthogonal
time-frequency resource units referred to as \emph{resource blocks}. 
We assume that the bandwidth and the duration of each resource block
are smaller than the coherence bandwidth and the coherence time of the
channel, respectively. This implies that the channel for each resource
block and each user can be considered to be frequency flat and
constant. More precisely, the channel from the transmitter of user $j$
(referred to as transmitter $j$) to the receiver of user $i$ (called
receiver $i$) on resource block $(t,f)$ is described by the channel
coefficient $h_{i,j}(t,f)\in \Complex$, which is a realization of some
stochastic process.
We assume that all resource blocks are statistically equivalent and
independent. Therefore, we can consider an arbitrary but fixed
resource block and drop the time and frequency index for simplicity.


Given a resource block, user $i\in\set N$ may experience interference
from other users $j\in \set N,j\neq i$. As a result, the performance
of user $i\in\set N$ depends in general on the vector $\vec h_{i}:=
\left(h_{i,1} , \ldots , h_{i,N}\right)^T\in \Complex^N$ of channel
coefficients $h_{i,j}\in\Complex$ from all transmitters $j\in\set N$
to receiver $i\in\set N$.
These channel vectors are grouped in the channel matrix $\vec H :=
\left(\vec h_{1}, \ldots, \vec h_{N} \right)$ which contains all
channel coefficients.

As discussed before, not all potential D2D users in $\set
N\setminus\set N_1$ need to be scheduled for transmissions. Therefore,
we define $\set S \subseteq \set N$ to be the index set of users
(cellular and D2D) scheduled for transmissions.  The signal observed
by receiver $i\in \set S$ is then
\begin{equation}
  \label{eq:inout}
  y_i = h_{i,i} s_i + \sum_{j\in \set S \setminus \{ i \}}
    h_{i,j} s_j + n_i, 
\end{equation}
where $s_j\in \Complex$ is the complex data symbol transmitted by node
$j$ and $n_i \sim \set C \set N(0 , \sigma^2_i)$ is additive noise at
receiver $i$. The transmitted data symbols are assumed to be
i.i.d. random variables with $\Ex{s_j}=0$ and $\Ex{|s_j|^2}=p_j$,
where the transmit power $p_j$ of user $j$ is assumed to be fixed
(i.e. we consider no power control). If user $i$ is scheduled for
transmission, then its achievable rate is assumed to be\footnote{Note
  that we could assume any strictly increasing function
  $f:\Reals_+\mapsto\Reals_+$ with $f(0)=0$ and
  $\lim_{x\to\infty}f(x)=+\infty$. }
\begin{equation}
\label{eq:rate}
r(\vec h_i , \set S) = \log\bigl(1+\SINR(\vec h_i , \set S)\bigr)
\end{equation}
where the SINR of receiver $i\in \set S$ is defined as the ratio of
the desired signal power to the sum of the interference and noise
power:
\begin{equation}
\label{eq:SINR}
\SINR(\vec h_i , \set S) := \frac{p_i|h_{i,i}|^2}{\sigma^2_i +
  \sum_{j\in \set S \setminus \{i \}}  p_j|h_{i,j}|^2 }.
\end{equation}
In what follows, we assume that each receiver $i$ has a rate (or
quality-of-service) requirement $\bar r_i$ and we define a feasible
scheduling decision as follows.
\begin{mydefinition}[Feasible scheduling decision] 
\label{def:feasible_scheduling}
Given a channel matrix $\vec H$, we say that a scheduling decision
$\set S$ is feasible if $\set N_1\subseteq\set S$ and $r(\vec h_i ,
\set S) \geq \bar r_i$ holds for each $i\in \set S\subseteq\set N.$
\end{mydefinition}
We emphasize that by the definition, $r(\vec h_i , \set S) \geq \bar
r_i$ for each $i\in\set N_1\subseteq\set S$ whenever $\set S$ is
feasible. In other words, the requirements of cellular users are
satisfied per definition and $\set N_1$ is a feasible scheduling
decision. As far as the potential D2D users in $\set N \setminus \set
N_1$ are concerned, the network controller may schedule them to be
paired with the transmissions in $\set N_1$, provided that (i) D2D
devices are in proximity to each other (see below) and (ii) the
resulting scheduling decision is feasible in the sense of
Def. \ref{def:feasible_scheduling}.

\subsection{D2D discovery and pairing with perfect CSI}
\label{sec:setup_perfect}

As mentioned in the introduction, two main steps towards establishing
a D2D communication are D2D discovery - also called proximity
discovery - and pairing. First we need to define the notion of
proximity.
\begin{mydefinition}[Proximity]
\label{def:proximity}
Given a channel realization, we say that two wireless devices are in
proximity to each other if the interference-free channel between them
is good enough to fulfill a given rate requirement.
\end{mydefinition}

In other words, proximity is necessary (but not sufficient) for
establishing a D2D link between two devices and D2D discovery is a
process of identifying \emph{D2D candidates} out of all potential D2D
users. Ideally, D2D discovery (and also pairing) should be based on
the achievable rates. If the network controller had namely perfect
knowledge of the channel matrix $\vec H$, it could compute the
achievable rates $r(\vec h_i , \set S),i\in\set S$, for all feasible
scheduling decision $\set S\subseteq\set N$. Thus, D2D discovery can
be performed as follows.

\begin{mydefinition}[D2D discovery with perfect CSI]
\label{def:discovery_perfect}
  Assuming that the network controller has perfect knowledge of $\vec
  h_i$ for some $i\in\set N\setminus\set N_1$, transmitter $i$ and
  receiver $i$ are said to be in proximity (to each other) if
  $i\in\set N_2$ where
\begin{equation}
  \label{eq:proximityIdeal}
  \set N_2 = \{i\in \set N \setminus \set N_1 : r(\vec h_i , \{ i \})
  \geq \bar r_i\}\subset\set N\,.  
\end{equation}
Therefore, $\set N_2$ is the set of all D2D candidates.
\end{mydefinition}

After performing D2D discovery, the network controller decides if D2D
candidates in $\set N_2$ are paired for transmissions with the
cellular users specified by $\set N_1$ to establish D2D links. The
optimal scheduling decision 
is found as follows. 

\begin{mydefinition}[Optimal pairing decision with perfect CSI]
\label{def:optimal_pairing}
  Under the assumption of perfect CSI at the network controller, an
  optimal scheduling decision $\set S \subseteq \set N_1 \cup \set
  N_2$ (that involves pairing decision) is a solution to
\begin{equation} 
\label{eq:pairingIdeal}
\begin{split}
& \max_{\set X \subseteq \set N_2} \sum_{i \in \set X \cup \set N_1}
r(\vec h_i , \set X \cup \set N_1) \\ 
& \text{subject to } r(\vec h_i , \set X \cup \set N_1) \geq \bar r_i
\text{ for all } i \in \set X \cup \set N_1\,.
\end{split}
\end{equation}
\end{mydefinition}

Since $\set N_1$ is assumed to be feasible decision scheduling, the
problem in (\ref{eq:pairingIdeal}) has always a solution in the sense
that if no D2D candidate can be paired with the cellular users, then
$\set S=\set N_1$ is the feasible scheduling decision. Note that since
$\set N_1$ is given, solving the pairing decision problem provides a
feasible scheduling decision $\set S$.


\section{Rate Estimation Based on Compressed Measurements}
\label{sec:rateEst}

One of the central tasks of the network controller is to perform
\emph{reliable} D2D discovery and pairing decisions. Here reliability
is to be understood in terms of the rate requirements of all users
which need to be satisfied permanently. In other words, the resulting
scheduling decisions $\set S$ must be feasible in accordance with
Def. \ref{def:feasible_scheduling} in spite of the lack of perfect
CSI. By Def. \ref{def:discovery_perfect} and Def.
\ref{def:optimal_pairing}, it is clear that reliable D2D discovery and
reliable pairing decisions require accurate estimates of the
achievable rates $r(\vec h_i , \set S)$ for any feasible scheduling
decision $\set S$. Therefore, accurate CSI is a crucial ingredient in
the design of reliable communication systems. 

In this section, we introduce a channel measurement and feedback
protocol together with different decoders that enables the central
controller to reliably estimate the achievable rates at relatively low
overhead costs. The measurement and rate estimation protocol is
summarized in Table
\ref{tab:protocol}. 

\begin{table}[htb]
 \caption{Measurement and rate estimation protocol.} 
 \begin{tabular}{|p{.2\linewidth}|p{.7\linewidth}|} 
   \hline
   Network controller & Transmit
   synchronization signal.\\ &\\ 
   Transmitters  &  Transmit sequences
   of $M$ pilot signals. \\ &\\ 
   Receivers & Measure superpositions of  pilot
   signals.\\ 
   & Quantize measurements and feed them back  to the network controller. \\ &\\ 
   Network controller & Estimate rates based on quantized compressed linear
   measurements. \\
   &Perform D2D discovery and make pairing/scheduling decision\\\hline
  \end{tabular}
 
  \label{tab:protocol}
\end{table}

\subsection{Random Channel Measurement}
\label{sec:random_measurements}

To reduce the signaling and coordination overhead for channel
measurements, all transmitters \emph{simultaneously} transmit $M\geq
1$ pilot signals.  In what follows, we use $\vec \phi_i\in
\Complex^{M}$ to denote the pilot signals sent by transmitter $i$,
which is the $i$th column of the so-called measurement matrix denoted
by $\vec \Phi = \left(\vec \phi_1 , \ldots , \vec \phi_N \right) \in
\Complex^{M\times N}$.  Then, according to \eqref{eq:inout}, the
vector of all $M$ signals observed by receiver $i$ can be written as
\begin{equation}
  \vec y_i = \vec \Phi \vec h_i + \vec n_i \in \Complex^M\; i\in\set N\,. \label{eq:y}
\end{equation}
Each receiver, say receiver $i\in\set N$, quantizes the vector of
channel measurements $\vec y_i$ using a quantization operator $\set Q:
\Complex^M \rightarrow \Complex^M$ and feed back the quantized values
to the network controller.  For simplicity, we make the following
assumption 
\begin{myassumption}
  \label{ass:quantization}
  We assume that $\set Q(\vec y_i) = \vec y_i + \vec {\bar n}_i$,
  where $\vec{ \bar n}_i$ is additive noise independent of $\vec y_i$.
  Furthermore, we assume an error and delay free feedback channel from
  all nodes to the network controller.
\end{myassumption}
By the assumption, the CSI at the network controller is
\begin{align}
  \vec z_i & = f(\vec y_i) + \vec \mu_i
   =  \vec \Phi \vec h_i + \vec \mu_i,   \label{eq:z}
\end{align}
where $\vec \mu_i:=\vec {\bar n}_i + \vec n_i$ is an additive noise
term that contains the measurement and quantization noise.  Further we
denote the matrix of all quantized channel measurements, which is
known to the network controller, by
$\vec Z := \left(\vec z_1,\ldots , \vec z_N \right)\in \Complex^{M
  \times N}$.



\subsection{Channel gain estimators}
\label{sec:system_estimators}

Given random channel measurements as described in the previous
subsection, the goal is to estimate CSI in the sense of minimizing the
gap between the achievable rates based on perfect CSI and their
estimates. To be precise, let $\vec z_i$ be compressed and quantized
CSI from receiver $i$ given by (\ref{eq:z}), and let $\beta(\vec z_i ,
j)$ be a deterministic function that estimates the channel gain
$|h_{i,j}|^2$. Hence,
\begin{equation}
\label{eq:function_beta}
|\hat  h_{i,j}|^2:=\beta(\vec z_i , j)\,,\quad i,j\in\set N\,,
\end{equation}
where
$\vec{\hat{h}_i}:=(\hat h_{i,1},\dotsc, \hat{h}_{i,N})\in\Complex^N$
is an estimate of $\vec{h}_{i}$ in the sense of
(\ref{eq:function_beta}). By (\ref{eq:rate}), the achievable rates are
proportional to the SINR, which in turn is a function of the channel
gains $|\vec h_{i,j}|^2$. As a result, it is sufficient to estimate
the channel gains instead of the complex channel coefficients.


In this paper, we consider different channel gain estimators specified
by the functions $\beta(\vec z_i , j)$. One class of function is given
by {\it channel gain estimation functions which are linear in the
  complex coefficients}:
\begin{mydefinition}[Linear channel gain estimator]
\label{def:LI}
Given the CSI $\vec z_i$ defined by \eqref{eq:z}, a linear channel
gain estimation function (for the channel coefficient $h_{i,j}$) is
given by
\begin{equation}
  \label{eq:linear}
  \beta_{l}(\vec z_i,j) = |\langle\vec \Psi \vec z_i  , \vec e_j\rangle|^2,
\end{equation} 
where the matrix $\vec \Psi \in \Complex^{N\times M}$ depends on the
measurement matrix $\vec \Phi$ and $\vec e_j$ is the $j$th column of
the identity matrix $\vec I_N$.
\end{mydefinition}
Another class of estimation functions is referred to as {\it
  non-linear channel gain estimation functions}:
\begin{mydefinition}[Non-linear channel gain estimator]
\label{def:CS}
Given the CSI $\vec z_i$ defined by \eqref{eq:z}, a non-linear channel
gain estimation function is given by
 \begin{equation*}
  \beta_{nl} (\vec z_i , j) = |\langle \alpha(\vec z_i) , \vec e_j\rangle|^2
 \end{equation*}
  where $\alpha :\Complex^M \rightarrow \Complex^N$ is some predefined
  non-linear function.
\end{mydefinition}

\subsection{Problem Statement: D2D discovery and pairing with imperfect CSI}
\label{sec:setup_imperfect}

The \emph{estimated achievable rate} $\hat{r}$ of user $i\in\set N$
can be seen as a function of $\vec{z}_i$. Therefore, given $\vec{h}_i$
and $\vec{z}_i$, the \emph{rate gap} of user $i$ depends on a
scheduling decision $\set{S}$, and is defined to be
\begin{equation}
\label{eq:def:Delta}
\Delta_i(\set{S}):=|\hat{r}(\vec{z}_i,\set{S}) - r(\vec{h}_i,\set{S})|\,,\;i\in\set{S}
\end{equation}
where the achievable rate $r$ is given by \eqref{eq:rate} and
\begin{equation}
  \label{eq:rateEstimFcn}
  \hat{r}(\vec{ z }_i,\set S) = \log\left(1 + \frac{ \beta(\vec
    z_i , i) p_i } {1 + \sum_{j\in \set S \setminus \{i \}} \beta(\vec
    z_i , j) p_j }\right).
\end{equation}
Here and hereafter $\beta(\vec{z}_i , i)$ is defined by
\eqref{eq:function_beta} and is the
estimated rate for given $\vec{z}_i$ and a scheduling decision
$\set S$.  For the ease of notation, in what follows, we write
$\Delta_i:=\Delta_i(\set S)$ if $\set S$ is clear from the context. We
use $\Delta_i(\{i\})$ as a basis for D2D discovery because it is the
rate gap of user $i\in\set N\setminus\set N_1$ in an interference-free
scenario. The rationale behind the definition of rate gap in
(\ref{eq:def:Delta}) comes from the rate requirements. In particular,
if we have $\Delta_i(\set{S})\leq\varepsilon$ for some known
$\varepsilon\geq 0$ and an arbitrary feasible $\set S$, then the
network controller is able to reliably perform D2D discovery and
pairing. 

To see this, let us first consider the problem of D2D discovery based
on compressed and quantized CSI $\vec{z}_i\in\Complex^M$. We assume
that the network controller can upper bound the rate gap such that
$\Delta_i(\{i\})\leq\varepsilon,i\in\set N\setminus\set N_1$ for some
$\varepsilon\geq 0$. It may be easily verified that, under this
assumption, the condition
$\hat{r}(\vec{z}_i , \{ i \}) \geq \bar r_i + \varepsilon$ implies
proximity so that $r(\vec{h}_i , \{ i \}) \geq \bar r_i$. As a result,
\begin{equation} 
\label{eq:PCSIdetection} 
\hat{\set N}_2 = \{i\in \set
  N\setminus \set N_1 : \hat{r}(\vec{z}_i , \{ i \}) \geq \bar r_i +
  \varepsilon \}\subseteq\set N_2
\end{equation}
is a set of device pairs that are in proximity to each other (see
Def. \ref{def:proximity}), and therefore are D2D candidates. So the
network controller is able to \emph{reliably} identify a subset of D2D
candidates, provided that it can upper bound the rate gap
$\Delta_i(\{i\})$. Notice that the cardinality $|\hat{\set N}_2|$ of
$\hat{\set N}_2$ is non-increasing in $\varepsilon$ and
$|\hat{\set N}_2|\to 0$ as $\epsilon\to\infty$. This means that
$\epsilon$ should be as small as possible to discover and identify as
many potential D2D users defined by (\ref{eq:proximityIdeal}) as D2D
candidates.  
In other words, we need a tight bound on each rate gap
$\Delta_i(\{i\}), i\in \set N$. 
Clearly, if $\varepsilon=0$, we have $\hat{\set N}_2=\set N_2$,
meaning that all potential D2D users have been discovered as D2D
candidates.

Having introduced the set $\hat{\set N}_2$, we are now in a position
to define optimal pairing decisions with imperfect CSI.
\begin{mydefinition}[Optimal pairing decisions with imperfect CSI]
\label{def:pairing_compressed}
For given $\hat{\set N}_2$ (with some $\varepsilon\geq 0$) and
$\vec{Z} = (\vec z_1 , \ldots ,\vec z_N)$ (compressed and quantized
CSI), we define an optimal scheduling decision
$\set{\hat S}=\set N_1\cup\set X\subseteq\set N_1\cup\hat{\set N}_2$
where $\set X\subseteq\set N_2$ is a solution to the following
problem:
\begin{align}
\label{eq:PCSIpairing}
& \set X:=\arg\max_{\set A \subseteq \set N_2} \sum_{i \in \set X \cup \set N_1}
\hat{r}(\vec{z}_i , \set A \cup \set N_1) \\ 
& \text{subject to } \hat{r}(\vec { z}_i , \set A \cup \set N_1) \geq
\bar r_i + \varepsilon
\text{ for all } i \in \set A \cup \set N_1
\end{align}
where $\hat{r}(\vec{ z }_i, \set S)$ is the estimated achievable rate
given by \eqref{eq:rateEstimFcn}.
\end{mydefinition}


\section{Rate gap analysis}
For different linear and non-linear
channel gain estimators we seek probabilistic bounds on the rate gap
$\Delta_i(\set S)$ of the form
\begin{equation}
  \label{eq:rateGap}
\Prob{\Delta_i(\set S) > d_i \,g(\xi , \varepsilon)}\leq  \varepsilon , 
\end{equation}
where $d_i > 0$ is a constant that depends on system parameters
(e.g. transmit powers, maximum number of scheduled users $|\set S|
\leq n$) and $g(\xi , \varepsilon)$ is a function of the measurement
and quantization noise. For simplicity we assume that the quantization
noise is bounded $\|\vec \mu_i\|_2 \leq \xi$. 

\subsection{Tail--Estimates for Subgaussian Random Matrices}
\label{subsec:concMeas}
The idea behind random pilots in channel probing is that if the amount
of (sufficiently) random signaling is above a certain threshold, the
response of channel is with high probability uniformly close to its
expectation.  This principle is used in various field of
high--dimensional geometry, such as random matrix theory and
compressed sensing. In fact, we proceed here along similar lines as in
\cite{Baraniuk2008} to prove RIP-properties based on
\emph{concentration inequalities} (see here also \cite{Jung2015} for
more details).

For an
in-depth treatment of this phenomenon, we refer the reader to
\cite{Ledoux2005}. A concise introduction can be found in
\cite{Tao2012}.  Throughout this section, we assume that the elements
of the measurement matrix $\vec \Phi$ are chosen at random and we
impose the following two conditions. 
\begin{myassumption}
\label{ass:com1}
The matrix is normalized such that for all $\vec a \in
\Complex^{N}$ 
\begin{equation*}
\Ex{\|\vec \Phi \vec a\|_2^2} = \|\vec a\|_2^2.
\end{equation*}
\end{myassumption}
\begin{myassumption}
\label{ass:com2}
For every  $\vec a \in \Complex^N$, the random variable
$\|\vec \Phi \vec a\|_2^2$ is strongly concentrated around its
expected value,
\begin{equation}
  \label{eq:concentration}
  \Prob{\bigl|\|\vec \Phi \vec a \|_2^2  - \|\vec a
    \|_2^2\bigr|   > \varepsilon \|\vec a \|_2^2}  \leq
  c_0e^{-\gamma(\varepsilon)}
\end{equation}
where $c_0 > 0$ is a constant, and $\gamma(\varepsilon)$ is a function
that depends on the distribution of $\vec \Phi$.
\end{myassumption}
Examples of measurement matrices that satisfy the concentration
inequality \eqref{eq:concentration} are matrices with rows that
are sub-Gaussian distributed isotropic vectors (see
e.g. \cite{Vershynin2010}). 
A real--valued random variable $X$ is called sub-Gaussian if there exists a constant $c>0$
such that the moment generating function is bounded from above by
\begin{equation}
  \label{eq:subGauss}
\Ex{\exp(Xt)} \leq \exp(c^2 t^2/2). 
\end{equation}
Examples of sub-Gaussian random variables are normally
distributed random variables and bounded random variables. 
In particular, if the elements of $\vec \Phi \in \Complex^{M\times N}$
are i.i.d. distributed according to $\phi_{i,j}\sim \set C\set N (0,1/M)$,
then $\Ex{\vec \Phi^H \vec \Phi}=\vec I_N$, and
$\Ex{\|\vec \Phi \vec a\|_2^2} = \vec a^H \Ex{\vec \Phi^H \vec \Phi}
\vec a = \| \vec a\|^2$. Moreover, it can be shown
(see e.g. \cite{Davenport2011}) that
\begin{equation}
  \label{eq:gaussPhi}
  \Prob{\bigl|\|\vec \Phi \vec a \|_2^2  - \|\vec a
    \|_2^2\bigr|   > \varepsilon \|\vec a \|_2^2}  \leq 
  2 \exp\left( \varepsilon^2 M \frac{\ln(2)-1}{2} \right).
\end{equation}
The sub-Gaussian assumption does not permit sufficiently structured
matrices $\vec \Phi$ but the result in \cite{Krahmer2011} shows that
RIP matrices with additional column randomization provide
Johnson--Lindenstrauss embeddings and this in turn implies a certain
concentration inequality of type \eqref{eq:concentration}.  We do not
further elaborate on this here, but refer the reader to \cite{Jung2015} for more details.


\subsection{Preliminary Result}
First, we introduce a general result that enables us to bound
$\Delta_i$ given by \eqref{eq:def:Delta} independent of the estimation function.  To simplify the notation we define the channel gain
$x_{i,j}:= |h_{i,j}|^2$, the vector of channel gains $\vec x_i :=
(x_{i,1},\ldots, x_{i,N})^T$ and the matrix of channel gains $\vec X:=
(\vec x_1,\ldots , \vec x_N)$. In a similar manner we define the
estimated channel gains as $\hat x_{i,j} := \beta(\vec z_i , j)$, the
vector of estimated channel gains $\vec{ \hat x}_i := (\hat
x_{i,1},\ldots, \hat x_{i,N})^T$ and the matrix of estimated channel
gains $\vec{\hat X}:= (\vec{\hat x}_1,\ldots , \vec{\hat x}_N)$.
\begin{mylemma} \label{lem:lipschitz} 
  Let the achievable rates $r(P, \set S,\vec h_i)$ be estimated by
  $\hat{r}_i(P,\set S, \vec z_i)$ defined in \eqref{eq:rateEstimFcn}.
  For any scheduling decision $\set S$, with $|\set S|\leq n$, and any
  channel gain estimation $\hat x_{i,j} = \beta(\vec z_i , j)$,
  \[
  \Delta_i(\set S) = |r_i(P, \set S,\vec h_i) - \hat{r}_i(P,\set S , \vec z_i)|
  \leq 2P \sum_{j\in \set S}|x_{i,j} - \hat x_{i,j}|,
  \]
  holds simultaneously for all $i\in \set S$.
\end{mylemma}
The proof is given in Section \ref{proof:lipschitz}.  
To control $\Delta_i$ it is sufficient to control $ \sum_{l \in \set N}
\left||h_{i,l}|^2 - |\hat h_{i,l}|^2\right|$ based on the measurements
$\vec z_i= \vec \Phi \vec h_i + \vec \mu_i$, defined in \eqref{eq:z}.
Hence, it is not necessary that we recover the vectors $\vec h_i$, for
all $i$. Instead, recovery of the vectors $\vec x_i$ is sufficient.
We stress that this is different from classical estimation theory (see
e.g. \cite{Luenberger1968}) where based on the measurements $\vec z_i$
minimization of the error $\| \vec h_i - \vec{\hat h_i}\|_2^2 =
\sum_{l \in \set N} |h_{i,l} - \hat h_{i,l}|^2 $ is considered.



\subsection{Non-Linear Rate Estimation}
In this subsection we study a non-linear channel gain estimation
function that uses concepts from compressed sensing to exploit the
structure of the channels.  More precisely, we assume that the channel
vectors are compressible, that is, for some $i\in \set N$ the channel
vector $\vec h_i$ is sparse or has at least fast decaying
magnitudes (after ordering). Compressibility of a given vector can be quantified by
decay order of
\[
\sigma_k(\vec x)_p :=  \min_{\vec{\hat x} \in \Sigma_k}\| \vec x - \vec{\hat x} \|_p,
\]
where $\Sigma_k:= \{ \vec x \in \Complex^N : |\text{supp}(\vec x)| \leq
k\}$ 
is the set of all $k$-sparse vectors.  The function
$\alpha$, defined in Definition \ref{def:CS}, is given by the solution
to the convex optimization problem
\begin{equation}\label{eq:l1NormMin} 
  \alpha(\vec z_i) = \underset{\vec x \in \Complex^{N}}{\arg \min} \| \vec x \|_{1}
  \quad \text{ subject to } \quad \| \vec \Phi \vec x - \vec z_i \|_2\leq
  \xi.
\end{equation}  
The parameter $\xi$ must be chosen such that $\|\vec \mu_i\|_2 \leq
\xi$. 
We will first review some basic results from compressed sensing and
then show how these results can be applied to obtain 
bounds on $\Delta_i$. Compressed sensing recovering
results can be divided in uniform and nonuniform recovery results. A
uniform recovery result means that one can recover all $k$-sparse
vectors -- with high probability -- from linear measurements with the
same matrix. Nonuniform recovery means that a fixed $k$-sparse vector
can be recovered with a randomly drawn measurement matrix, with high
probability. Uniform recovery results are obviously stronger since
they imply nonuniform recovery. To streamline the presentation we
consider only uniform recovery.

One class of uniform recovery results are based on the restricted
isometry property (RIP) (see e.g. \cite{Foucart2013})  of the measurement matrix $\vec \Phi$.  The
RIP is defined as follows.
\begin{mydefinition}
  An $M\times N$ matrix $\vec \Phi$ satisfies the RIP of order $k\geq
  1$, if there exists $ 0 \leq \delta_k$ such that the
  inequality
  \[ (1-\delta_k)\|\vec x \|_2^2 \leq \|\vec \Phi \vec x \|_2^2 \leq
  (1 + \delta_k)\|\vec x \|_2^2,
  \] holds for all $\vec x\in \Sigma_k$. The smallest number $\delta_k = \delta_k(\vec
  \Phi)$ is called the restricted isometry constant of the matrix
  $\vec \Phi$.
\end{mydefinition}
Many ensembles of random matrices are known to satisfy the RIP with
high probability.  An important class of random matrices are matrices
with elements that are i.i.d. sub-Gaussian distributed.  In
particular, if $X\sim \set C \set N(0,\sigma^2)$, then $\Ex{\exp(Xt)}
\leq \exp(\sigma^2t^2/2)$ and therefore, according to
\eqref{eq:subGauss}, $X$ is sub-Gaussian. 

\begin{figure}[tb]
  \centering
  \includegraphics[width=.75\linewidth]{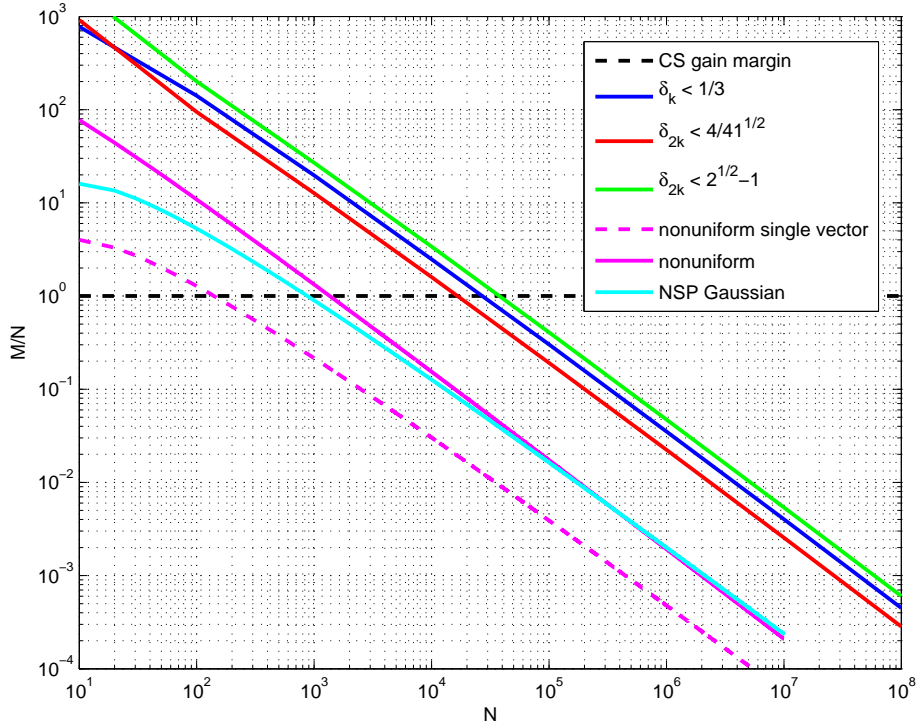}
  \caption{Bounds on compression ratio $M/N$ over system size
    $N$. Maximal compression to achieve perfect reconstruction with
    probability $ \varepsilon = 0.9$ fixed sparsity $k=10$.}
  \label{fig:CSgain}
\end{figure}

For concreteness we assume that the elements of $\vec \Phi$ are
distributed complex Gaussian $\phi_{i,j} \sim \set C \set N
(0,1/M)$. In fact, this assumption enables us to explicitly compute
most of the constants that would otherwise depend on the distribution
of $\vec \Phi$. We stress that more general results for sub-Gaussian
measurement matrices can be found for example in
\cite{Eldar2012,Foucart2013} and references therein. The following
theorem which is proved in \cite[Theorem 9.27]{Foucart2013} enables us
to bound the RIP constant of $\vec \Phi$. To be self contained, we
state the theorem in our notation.
\begin{mytheorem}[{\cite[Theorem 9.27]{Foucart2013}}]\label{theo:RIP}
  Let $\vec \Phi$ be a random $M\times N$ matrix with i.i.d. elements
  distributed according to $\phi_{i,j} \sim \set C \set N
  (0,1/M)$. Assume that 
\[ M \geq 2\eta^{-2} \left(k \ln(e N / k) +
  \ln(2\varepsilon^{-1} )\right),
  \]
  with $\eta , \varepsilon \in (0,1)$.
  Then the RIP constant $\delta_k$ of $\vec \Phi$ satisfies
  \[
  \delta_k \leq 2 \left( 1 + \frac{1}{\sqrt{2 \ln(e N /k)}}  \right)\eta
  + \left( 1 + \frac{1}{\sqrt{2 \ln(e N /k)}}  \right)^2 \eta^2,
  \]
  with probability $1 - \varepsilon$.
\end{mytheorem}
As pointed out by \cite[Remark 9.28]{Foucart2013} the statement of the
last theorem can be simplified by using  
$\delta_k \leq \delta \leq C_1 \eta$ with $C_1 = 2(1 + \sqrt{1/2}) + (1 +
\sqrt{1/2})^2$ such that
$
M \geq  2C_1^2 \delta^{-2} \left(k \ln(e N / k) +  \ln(2\varepsilon^{-1} )\right)
$ 
yields $\delta_k \leq \delta$. 
According to Lemma \ref{lem:lipschitz} we can control the rate
gap $\Delta_i$ by controlling $\| \vec x_i
- \vec {\hat x}_i \|_2$. If the
measurement matrix satisfies the RIP of order $k$ with $\delta_k <
1/3$, the following theorem provides an error estimate.
\begin{mytheorem}[ {\cite[Theorem 3.3]{Cai2013}}] \label{theo:CSgeneral}
  Suppose $\vec \Phi$ satisfies the RIP of order $k$ with $\delta_k <
  1/3$. Let the measurements be given by $\vec z = \vec \Phi \vec h +
  \vec \mu $, according to \eqref{eq:z}, with $\|\vec \mu \|_2 \leq
  \xi$.  Then for any $\vec h\in \Complex^N$ the solution $\vec{\hat
    h} = \alpha(\vec z)$ to \eqref{eq:l1NormMin} obeys
  \begin{equation}
    \label{eq:CSerror}
    \|\vec h - \vec{ \hat h} \|_2 \leq 
    C_2(\delta_k) \frac{\sigma_k(\vec h_i)_1}{\sqrt{k}} 
    + 2 C_3(\delta_k)\xi,
  \end{equation}
  where
  $C_2(\delta)=\frac{2\sqrt{2}(2\delta+\sqrt{(1-3\delta)\delta})+2(1-3\delta)}{1-3\delta}$
  and $C_3(\delta) = \frac{\sqrt{2(1+\delta)}}{1-3\delta}$ are
  constants.
\end{mytheorem}
The theorem is proved in \cite[Theorem 3.3]{Cai2013}.  We stress that
many similar error bounds for Problem \ref{eq:l1NormMin} and related
problems are known. The probably most popular error bound was provided
in the seminal paper \cite{candes2006stable}, which requires that the
measurement matrix has a RIP constant $\delta_{2k} \leq \sqrt{2} -
1$. A better error bound is given in \cite[Theorem 6.12]{Foucart2013}
where $\delta_{2k} \leq 4 / \sqrt{41}$ is required. Recently
\cite{Cai2014} showed that $\delta_{2k} <
  1/\sqrt{2}$ is sufficient.
Figure \ref{fig:CSgain} depicts the system size $N$ over the
compression ratio $M/N$ for different RIP constants.  The number of
measurements is evaluated according to Theorem \ref{theo:RIP}.  To
obtain a significantly reduced number of measurements, the number of
links $N$ must be large.  Figure \ref{fig:CSgain} includes
also bounds on the number of measurements for non-uniform
recovery. Non-uniform recovery results provide error bounds for much
smaller system sizes $N$. However, we stress that the RIP is only a
sufficient condition for recovery.
From Theorem \ref{theo:RIP}, Theorem \ref{theo:CSgeneral} and Lemma
\ref{lem:lipschitz} we devise the following corollary.
\begin{mycorollary}\label{cor:CS}
  Let $\vec \Phi$ be a random $M\times N$ matrix with i.i.d. elements
  distributed according to $\phi_{i,j} \sim \set C \set N
  (0,1/M)$. Suppose the measurements are given by $\vec z_i = \vec \Phi \vec
  h_i + \vec \mu_i $, according to \eqref{eq:z}, with $\|\vec \mu \|_2
  \leq \xi$.  If 
  \[
  M \geq  2C_1^2 \delta^{-2} \left(k \ln(e N / k) +
    \ln(2\varepsilon^{-1} )\right), 
  \]
  with $\delta \leq \delta_{2k}<1/3$ and $\|\vec h_{i} \|_2 \leq a_i$, then for all $\{\vec h_i\}_{i\in\set
    N}$ the solutions $\{\vec{\hat h}_{i}\}_{i\in\set N}$ to
  \eqref{eq:l1NormMin} obey
  \[
  \Prob{\exists i\in \set N : \Delta_i > 2P q(\vec h_i , \xi) (2a_i + q(\vec h_i , \xi))} \leq 
  \varepsilon,
  \]
  with $q(\vec h_i , \xi) = C_2(\delta_k) \frac{\sigma_k(\vec h_i)_1}{\sqrt{k}} 
    + 2 C_3(\delta_k)\xi$ and  $C_2(\delta),C_3(\delta)>0$ as in Theorem \ref{theo:CSgeneral}.
\end{mycorollary}
The proof is given in Section \ref{sec:proof:cor:CS}.  We point out
that, if the number of measurements are in the order of $\set O(k
\ln(e N / k)$ and, for all $i\in \set N$, the channels $\vec h_i$ are
$k$-sparse (i.e. $\sigma_k(\vec h_i)_1 = 0$), then the rate estimation
error $\Delta_i$ remains bounded. Moreover, in the noiseless case
($\xi = 0$) perfect recovery can be achieved. However, for both cases
the system size $N$ must be sufficiently large as said before and
illustrated in Figure \ref{fig:CSgain}.


\subsection{Linear Rate Estimation} 
In this subsection we derive bounds on the rate gap $\Delta_i$ for
linear channel gain estimation functions. First, we prove a general
theorem that is valid for any linear estimation function defined in
Definition \ref{def:LI} and any ensemble of measurement matrices that
satisfies the concentration of measure inequality
\eqref{eq:concentration}.  We have the following general result, which
is the main result in this chapter.
\begin{mytheorem}\label{theo:RateEst}
  Let channel state information be given by any linear estimation
  function $\beta(\vec z_i,j)= |\langle \vec \Psi \vec z_i , \vec
  e_j\rangle|^2$, with $\vec \Psi = \vec \Phi^H \vec A$ where $\vec A$
  is a positive semi-definite matrix. If $\vec \Phi$ fulfills the
  concentration inequality \eqref{eq:concentration} and the number of
  active transmissions is bounded by $1\leq |\set S| \leq n$, then for
  any fixed channels $\vec H = (\vec h_1 , \ldots, \vec h_N)$ and any $u_0\geq 0$, $\rho_0\geq 0$
  and $\varepsilon \geq 0$,
\begin{multline}
    \Prob{\exists \set S \subset \set N , |\set S| = n, \exists i \in \set S  :\Delta_i(\set S) > 2P \|\vec h_i \|_2^2 (4
      \sqrt{n}(1+u_0)\varepsilon + \rho_0)} \\ \leq \exp(\log(4n^2) + n \log(Ne/n) - \gamma(\varepsilon)) 
    +\exp(n \log(Ne/n)) \Prob{s_{\max}(\vec \Psi \vec \Phi) >
      u_0 } \\ + \exp(n \log(Ne/n) )  \max_{i\in \set N}\Prob{\|\vec \Psi \vec{\bar \mu}_i \|_2(\|\vec
      \Psi \vec{\bar \mu}_i\|_2 + 2\|\vec \Psi\vec \Phi\vec{\bar h}_i
      \|_2) > \rho_0},
 \end{multline}
\end{mytheorem}
The proof is deferred to Section \ref{sec:proofRateEst}. Clearly the
bound depends on the choice of $\vec \Psi$ and the distribution of
$\vec \Phi$. The latter determines the function $\gamma(\varepsilon)$.
However the theorem is rather general and enables the evaluation of
different linear estimation functions under different assumptions on the
channels and under different distributions of the measurement matrix $\vec
\Phi$.

To illustrate the strength of Theorem \ref{theo:RateEst} let us assume
that the channel vectors are $k$-sparse, $\vec h_i\in \Sigma_k$ for
all $i$, and consider the following estimation function and
measurement matrix. Let the elements of $\vec \Phi$ be distributed
complex Gaussian and define the linear estimation function as
\begin{equation}
  \label{eq:l2_decoder}
  \beta_l(\vec z_i , j) = |\langle \vec \Phi^+\vec z_i, \vec e_j \rangle|^2, 
\end{equation}
where $\vec \Phi^+$ is defined as the pseudo inverse $\vec \Phi^+ =
\vec \Phi^H(\vec \Phi \vec \Phi^H)^{-1},$ for $M<N$. We devise the
following corollary.
\begin{mycorollary}\label{coro:linEst} Under the assumptions of Theorem
  \ref{theo:RateEst}. Let $M<N$. Suppose that the elements of $\vec
  \Phi$ are distributed $\phi_{i,j} \sim \set C \set N (0,1/M)$.  Let
  $\beta_l(\vec z_i , j) = |\langle \vec \Phi^+\vec z_i, \vec e_j
  \rangle|^2$. Assume that $\|\vec e\|_2 = 0$ and for all $i\in \set
  N$ we have $\vec h_i\in \Sigma_k$ and $\vec h_{i,j} \sim \set C
  \set N (0,1)$ for all $j \in \text{supp}(\vec h_i)$. We have
  \[ \Prob{\exists i\in \set N : \Delta_i > 16 P\sqrt{\frac{\kappa
        n}{M}} \left(\sqrt{2}\ln\left(\frac{4 nN\left( \multi{N}{n} \right) +
          1}{\varepsilon}\right) + k \sqrt{\ln\left(\frac{4 nN\left( \multi{N}{n} \right) +
            1}{\varepsilon}\right)} \right)} \leq \varepsilon,
  \]
  with $\kappa = 2/(1-\log(2))$.
\end{mycorollary}
The proof is given in Section \ref{proof:linEst}. A few remarks are in
place. For fixed transmit powers $P$, a fixed system size $N$, a given
error probability $\varepsilon$ and a fixed number of active links
$n$, the rate estimation error scales with $\sqrt{1/M}$, which is also
in accordance with the estimation results in \cite[Theorem
4.1]{davenport2006detection}, where essentially the same scaling is
achieved. As was expected, the linear decoding function is not able
to achieve perfect recovery (for $M<N$). Perfect recovery can only be
achieved by the compressed sensing based decoder but comes at the cost
of additional complexity. However, the simulations in the next section
show that the linear decoder performs reasonably well when applied to a small
systems. Moreover, a linear decoder can be used to perform a subset
selection and to reduce the problem size for non-linear algorithms.

\section{Numerical Examples}\label{sec:sim}

We consider a cellular system with one base station and $25$
users. Every node has a single antenna.  The users are grouped in $G$
user groups $\set G_g$, $g=1,2,\ldots,G$. Users within the same user group
experience the same path loss. The channels from users in $i\in \set G_f$ to users
$j \in \set G_g$ are given by
\begin{equation}
  \label{eq:channelModel}
h_{j,i} = a_{g,f} b_{j,i} \in \Complex,  
\end{equation}
where $b_{j,i}\sim \set C \set N(0,1)$ denotes the small scale fading
coefficient and $a_{g,f}$ denotes the distance dependent path loss
coefficient, with $a_{g,g}=1$ for all $g$. A similar channel model was
used in \cite{Huh2010} to model large cellular networks with
co-located users.  Under certain assumptions the channel matrix $\vec H$
is compressible. More precisely, the matrix $\vec H$ can be
approximated by a low rank and/or sparse matrix $\vec{\hat H}$, if the
user groups $\set G_g$ are of sufficient size and/or the path loss
coefficients $a_{r,g}$ decay sufficiently fast.

 We compare two setups: i) $5$
groups of $5$ users each, the path loss coefficients are chosen as
$10^{z/10}$, with $z$ uniformly distributed in $[0,1]$.  ii) $25$
users all in the same group and path loss coefficient is $a_{i,g}=1$ for
all $i,g$, i.e., all channels are i.i.d. complex Gaussian distributed.
The rate requirement is set to $\bar r = 1/10 \log(1 + P)$.  Problem
\ref{eq:l1NormMin} was solved using the Tfocs toolbox
\cite{becker2011tfocs}.

\begin{figure}[tb]
  \centering
  \includegraphics[width=.75\linewidth]{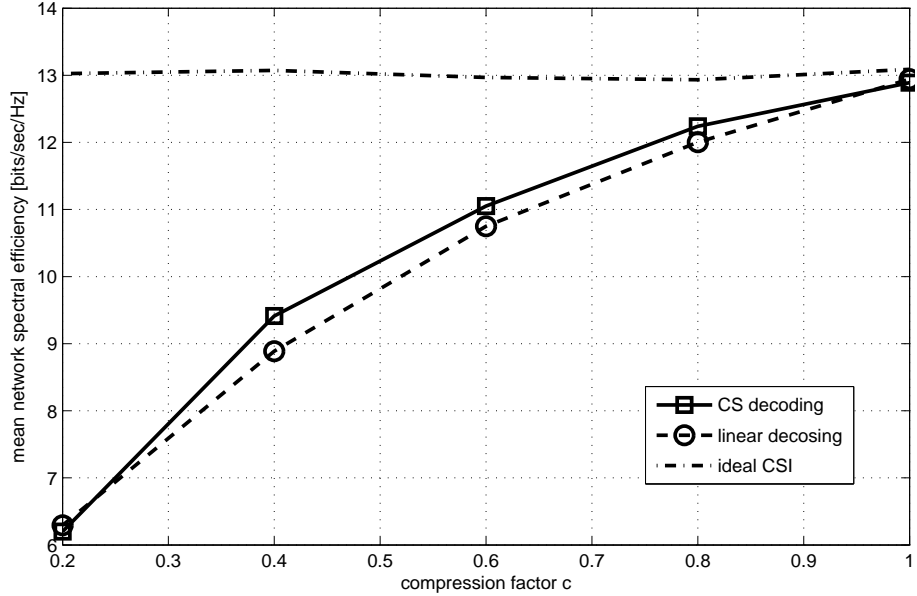}
  \caption{Average sum-rate over compression factor $M/N$; Setup: $25$
    users, $1$ base station, perfect feedback channel (no feedback and
    quantization noise), single group; channel matrix i.i.d. Gauss and
    not compressible. Comparison of linear and non-linear rate
    estimation.}
  \label{fig:A}
\end{figure}

\begin{figure}[tb]
  \centering
   \includegraphics[width=.75\linewidth]{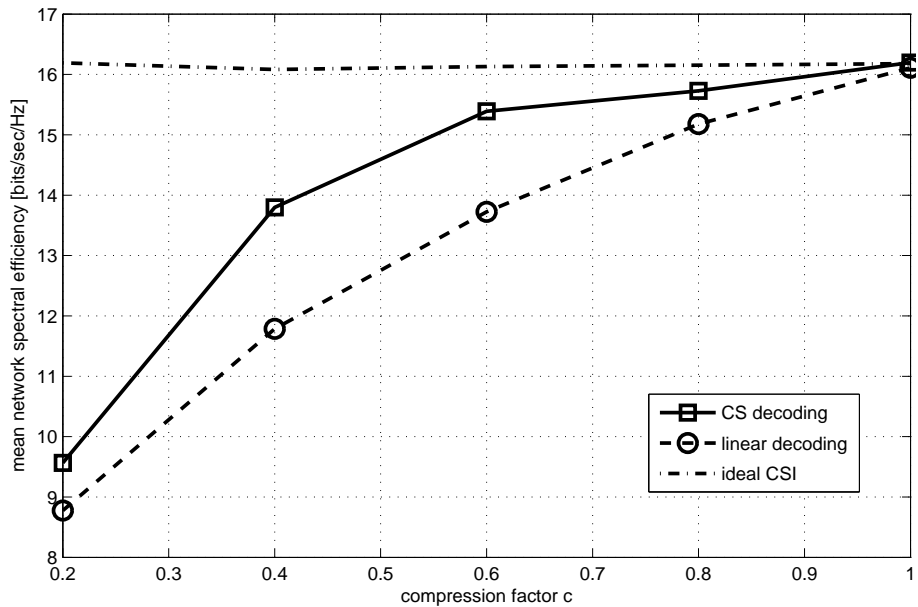}
  \caption{Average sum-rate over compression factor $M/N$; Setup: $25$ users, $1$ base station, perfect feedback
    channel (no feedback and quantization noise), 5 group of 5 users each;
    channel matrix compressible, single group; channel matrix i.i.d. Gauss and
    not compressible. Comparison of linear and non-linear rate
    estimation.}
  \label{fig:B}
\end{figure}

We compare the solution to problem \eqref{eq:PCSIpairing} for the
non-linear compressed sensing estimation function \eqref{eq:l1NormMin}
and the linear estimation function \eqref{eq:l2_decoder}. In the
simulations $\varepsilon = 0$, since the analytic results do not give
tight bounds for systems with $N=25$. Nevertheless the results in
Figure \ref{fig:A} show that linear estimation performs very close to
the much more complex compressed sensing based estimation. Figure
\ref{fig:B} shows that if the channel matrix is compressible the
compressed sensing estimation function performs better than
the linear estimation function. Since the considered systems are rather
small it can be expected that the gain of compressed sensing increases for
larger systems.

 \section{Conclusion}
\label{sec:Conclusion}
We developed a channel sensing and reconstruction protocol that
enables the network controller to estimate the achievable rates based
on compressed non-adaptive measurements. The scaling of the estimation
error at the network controller has been analyzed for linear and
non-linear decoding functions. Scaling results for the non-linear
decoding function where shown to follow from well known compressed
sensing results. However, for a small to moderate system size $N$ the
compressed sensing results do not provide reasonable performance
bounds.  For linear decoding functions we derived a general result
which can be used to analyze the performance of a variety of linear
decoding functions and measurement matrices. For a linear decoding
function based on the pseudo inverse and Gaussian measurement matrices
we investigated the scaling of the rate estimation error with the
number of measurements.

The measurement protocol is based on a few simplifications which
render the direct application in practical systems rather difficult.
For example, the assumption of perfect time and frequency
synchronization is hard (if not impossible) to achieve in distributed
networks with a huge number of devices. To this end, the analog coding
developed in \cite{Goldenbaum:Stanczak:11a} can be used to relax the
requirements on the synchronization.

Future work may also include the exploration of different linear and
non-linear decoding functions. To this end, Theorem \ref{theo:RateEst}
provides a good basis to evaluate different linear decoding functions.
For non-linear decoding functions applications of matrix recovery and
other compressed sensing related approaches are a promising research
direction.  Extensions to other network architectures are another
prospective direction. Coordinated transmission techniques where
groups of devices (or antennas) are jointly transmitting with
beamforming vectors $\vec w$ given by some finite codebook can be
analyzed with the proposed framework by estimating $|\langle \vec h_i , \vec
w\rangle|$. 
\newpage
\section{Proofs}

\subsection{Proof of Lemma \ref{lem:lipschitz}}
\label{proof:lipschitz}
\begin{IEEEproof} 
  For each $i$ the corresponding rate gap $\Delta_i$
  can rewritten using the abbreviations
  $L(s):=\log(1+s)$, $q_{j}: = p_j|h_{i,j}|^2$ 
  and $\hat q_{j}: = p_j|\hat h_{i,j}|^2$:
  \begin{equation}\begin{split}
      \Delta_i & = \left| L\left(\frac{q_{j}}{1 +
            \sum_{l\in \set S \setminus \{i \}} q_{l} }\right)
        - L\left(\frac{\hat q_{j}}{1 + \sum_{l\in \set S
              \setminus \{i \}} \hat q_{l} }\right) \right|\\
      & = \left| \log\left( \frac{1 + \sum_{l\in \set S}
            q_{l}}{1 + \sum_{l\in \set S} \hat q_{l} }\right)
        + \log\left(\frac{1 + \sum_{l\in \set S \setminus \{i \}}
            \hat q_{l} }{1 + \sum_{l\in \set S \setminus \{i \}}
            q_{l}}\right)  \right|\\
      & = \left| L\left( \frac{\sum_{l\in \set S} q_{l}
            - \hat q_{l}}{1 + \sum_{l\in \set S} \hat q_{l}
          }\right) + L\left(\frac{\sum_{l\in \set S \setminus
              \{i \}} \hat q_{l} - q_{l}}{1 + \sum_{l\in
              \set S \setminus
              \{i \}} q_{l}}\right) \right| \\
      & \leq \Biggl| L\left( \frac{\sum_{l\in \set S}
          q_{l} - \hat q_{l}}{1 + \sum_{l\in \set S} \hat
          q_{l} }\right)\Biggr| + \Biggl| L\left(
        \frac{\sum_{l\in \set S \setminus \{i \}} \hat q_{l} -
          q_{l} }{1 + \sum_{l\in \set S \setminus
            \{i \}} q_{l}}\right) \Biggr| \\
      & \leq 2L\left( \sum_{l\in \set S} |q_{l} - \hat
        q_{l}|\right),
    \end{split}
  \end{equation}
  where the first inequality follows from the triangle inequality and
  the second inequality follows from Jensen's inequality and the fact
  that the denominators are positive.
  Since, $L(x) \leq x$ for $x\geq 0$ and
  by assumption $p_j\leq P$, for all $j$, we obtain the first claim
  \[
  \Delta_i  \leq 2P \sum_{l\in \set S} ||h_{i,l}|^2 - |\hat h_{i,l}|^2|.
  \]
\end{IEEEproof}

\subsection{Proof of Corollary \ref{cor:CS}}
\label{sec:proof:cor:CS}
\begin{IEEEproof}
  Using Lemma \ref{lem:lipschitz}, the Cauchy-Schwarz inequality and
  the reverse triangle inequality we get
  \begin{align}
    \Delta_i &\leq  2P\sum_{j\in \set
      S}|x_{i,j} - \hat x_{i,j}| \\
    &\leq 2P\sum_{j\in \set
      N}|(|h_{i,j}| - |\hat h_{i,j}|)(|h_{i,j}| + |\hat h_{i,j}|)| \\ 
    &\leq 2P \left\| \vec h_{i} - \vec{\hat h}_{i}\right\|_2
    \left\||\vec h_{i}| + |\vec{\hat h}_{i}|\right\|_2\\
    & \leq 2P \left\| \vec h_{i} - \vec{\hat h}_{i}|\right\|_2\left(  2\|
      \vec h_i\|_2 +  \left\| \vec h_{i} - \vec{\hat h_{i}} \right\|_2
    \right). \label{theEND}
  \end{align}
  By assumption $M \geq 2C_1^2 \delta^{-2} \left(k \ln(e N / k) +
    \ln(2\varepsilon^{-1} )\right)$, with $\delta < 1/3$, such that
  $\vec \Phi$ satisfies the RIP with probability at least $1 -
  \varepsilon$.  Hence, we can use Theorem \ref{theo:CSgeneral}
  and plug \eqref{eq:CSerror} in \eqref{theEND}.  Finally, defining $q(\vec h_i , \xi) = C_2(\delta_k) \frac{\sigma_k(\vec h_i)_1}{\sqrt{k}} 
  + 2 C_3(\delta_k)\xi$ the claim follows. 
\end{IEEEproof}

\subsection{Proof of Theorem \ref{theo:RateEst}}
\label{sec:proofRateEst}
The prove of Theorem \ref{theo:RateEst} is developed in several steps.
\begin{mylemma}\label{lem:prob}
  Let $X$ and $Y$ be two non-negative real random variables. If $f:\Reals
  \times \Reals \rightarrow \Reals$ is monotonically increasing in the
  second input and $y_0>0$ is
  a positive constant, then
\[
\Prob{f(X,Y) > \varepsilon} \leq \min_{y_0\geq 0}\Prob{f(X,y_0) > \varepsilon} + \Prob{Y>y_0}.
\]
\end{mylemma}
\begin{IEEEproof}
  First assume that the random variable $Y$ is bounded by $Y>y_0$. In
  this case the claim is trivially true, since
  $\Prob{Y>y_0}=1$. Therefore, assume that $\Prob{Y\leq
    y_0} > 0$. 
  We will abbreviate $Z=f(X,Y)$ and $Z_0=f(X,y_0)$.
  For any arbitrary but fixed $y_0\geq 0$ we have,
  \begin{equation}\begin{split}
    &\Prob{Z > \varepsilon} = 1 - \Prob{Z \leq \varepsilon|Y \leq y_0} - \Prob{Z \leq  \varepsilon | Y >  y_0}\\
    & \leq 1 - \Prob{Z \leq \varepsilon | Y \leq
      y_0} 
    \leq 1 - \Prob{Z_0 \leq \varepsilon | Y \leq
      y_0} \\
    & = 1 - \frac{\Prob{\{Z_0 \leq \varepsilon\} \cap \{Y \leq
        y_0\}}}{\Prob{Y\leq y_0}} \\
    & = 1 - \frac{1 - \Prob{\{Z_0 > \varepsilon\} \cup \{Y >
        y_0\}}}{\Prob{Y\leq y_0}}\\
    & \leq 1 - \frac{1 - \Prob{Z_9 > \varepsilon} - \Prob{Y >
        y_0}}{\Prob{Y\leq y_0}} \\
    & \leq \Prob{Z_0 > \varepsilon} +\Prob{Y > y_0},
  \end{split}\end{equation}
where we first used De Morgan's law and then the union bound. 
\end{IEEEproof}

\begin{mylemma} 
  \label{lem:motiNOISE}
  Let $\set V = \{\vec v_1,\ldots,\vec v_n \} \subset \sphere^{N-1}$
  be an arbitrary but fixed set of mutually orthogonal vectors ($n\leq
  N$), $\vec
  \Psi = \vec \Phi^H \vec A \in \Complex^{N \times M}$ and $\vec A \in
  \Complex^{M\times M}$ be a positive semi-definite matrix. If $\vec w = \vec
  \Phi \vec u + \vec e$ and $\vec \Phi$ is a
  $M\times N$ random matrix that is isotropically distributed and satisfies the concentration
  inequality \eqref{eq:concentration}, 
  then for any fixed $\vec u \in\sphere ^{N-1}$ and any fixed $\vec e \in \Complex ^M$
  \begin{equation}\begin{split}
      \lefteqn{\Prob{\left|\sum_{i=1}^n |\langle \vec u , \vec v_i\rangle|^2 -
        |\langle \vec \Psi \vec w , \vec
        v_i\rangle |^2 \right| > 4\sqrt{n}(1+u_0)\varepsilon + \rho_0 }}  \\        
      &\leq 4n  \exp(-\gamma\left(\varepsilon\right)) +
      \Prob{s_{\max}(\vec \Psi \vec \Phi)  > u_0 } \\
      & \quad +\Prob{ \|\vec \Psi \vec e \|_2 ( \|\vec \Psi \vec e \|_2
        + 2\| \vec \Psi \vec \Phi \vec u \|_2) > \rho_0} 
    \end{split}
  \end{equation}
  holds, where $\gamma(\varepsilon)$ depends on the distribution
  of $\vec \Phi$ and $\rho_0, u_0\geq 0$ are  positive constants.
\end{mylemma} 
\begin{IEEEproof}
  Consider the vectors $\vec a,\vec b,\vec c\in\mathbb{C}^n$ with
  elements $a_i=\langle \vec{u} , \vec v_i\rangle$, $b_i=\langle \vec
  \Psi \vec \Phi \vec{u},\vec v_i\rangle$ and $c_i=\langle \vec \Psi
  \vec e,\vec v_i\rangle$.  Obviously $\lVert \vec a\rVert_2\leq1$,
  $\lVert \vec b\rVert_2\leq\lVert \vec \Psi \vec \Phi
  \vec{u}\rVert_2$ and $\lVert \vec c\rVert_2\leq\lVert \vec \Psi \vec
  e\rVert_2$.
  \begin{equation}\begin{split}
      D:=&\sum_{i=1}^n \left||a_i|^2 - |b_i+c_i|^2 \right|
      =\sum_{i=1}^n \left||a_i|^2
        - |b_i|^2 - |c_i|^2 - 2 \Re
        \left( b_i \bar c_i\right)\right|\\
      & \leq \sum_{i=1}^n \left||a_i|^2 - |b_i |^2\right| + |c_i|^2 + 2 |b_i \bar c_i|\\
      & \leq \sum_{i=1}^n \left||a_i|^2 - |b_i |^2\right|+
      \lVert \vec c\rVert(1+2\lVert \vec b\rVert)\\  
      & = \sum_{i=1}^n \left|(|a_i| - |b_i |)(|a_i| + |b_i |)\right| +  \lVert \vec c\rVert_2(1+2\lVert \vec b\rVert_2)\\
      &\leq \lVert |\vec a| - |\vec b|\rVert_p\cdot\lVert |\vec a| + |\vec b|\rVert_q+\lVert \vec c\rVert_2(1+2\lVert \vec b\rVert_2)\\
      &\leq \lVert \vec a - \vec b\rVert_p\cdot(\lVert \vec a\rVert_q + \lVert \vec b\rVert_q)+\lVert \vec c\rVert_2(1+2\lVert \vec b\rVert_2)
    \end{split}
  \end{equation}  
  Recall, that $\vec b$ and $\vec c$ are random vectors. We 
  apply now Lemma \ref{lem:prob} twice. First, for the non--negative random variables 
  $X=\lVert \vec a - \vec b\rVert_p\cdot(\lVert \vec a\rVert_q +
  \lVert \vec b\rVert_q)$ and $Y=\lVert \vec c\rVert_2(1+2\lVert \vec
  b\rVert_2)$, for any $0\leq \rho_0$, we have,
  \begin{equation}
    \begin{split}
      \Prob{D > \varepsilon'}
      \leq &
      \underbrace{\Prob{\lVert \vec a - \vec b\rVert_p\cdot(\lVert \vec a\rVert_q + \lVert \vec b\rVert_q)+\rho_0>\varepsilon'}}_{(i)} + 
      \Prob{ \lVert \vec c\rVert(1+2\lVert \vec b\rVert) >
        \rho_0}. 
      \label{eq:CS:step1}
    \end{split}
  \end{equation}  
  Second, for $X=\lVert \vec a-\vec b\rVert_p$
  and $Y=\lVert \vec a\rVert_q + \lVert \vec b\rVert_q$, for any $y_0>0$, we have,
  \begin{equation}
    \begin{split}
      \text{(i)}
      &=\Prob{\lVert \vec a - \vec b\rVert_p\cdot
        (\lVert \vec a\rVert_q + \lVert \vec b\rVert_q)>\varepsilon'-\rho_0}\\
      &\leq\Prob{\lVert \vec a - \vec b\rVert_p>\frac{\varepsilon'-\rho_0}{y_0}}
      +\Prob{\lVert \vec a\rVert_q + \lVert \vec b\rVert_q\geq y_0}
    \end{split}
  \end{equation}       
  By assumption $\vec \Psi = \vec \Phi^H \vec A = \vec \Phi^H \vec A^{1/2}\vec
  A^{1/2}$, where $\vec A^{1/2}$ is the principal square root of $\vec
  A$. From the definition of $\vec a$ and $\vec b$ it follows that
  \begin{align}
    \lVert \vec a-\vec b\rVert_p
    =\lVert\{| \langle \vec{u},\vec v_i \rangle-\langle \vec B \vec{u},\vec B\vec v_i \rangle|\}_{i=1}^n\rVert_p
    \leq\lVert\vec m\rVert_p,
    \label{eq:lastMax}
  \end{align}
  where the $n$ components $m_i$ of the vector $\vec m$ follow from
  the polarization identity as,
  \begin{equation}\begin{split}
      |a_i-b_i|
      &=  \frac{1}{4}\Bigl|\sum_{\xi\in\{\pm 1,\pm i\}} \xi( 
      \|\vec{u} + \xi\vec v_i \|_2^2  - \|\vec B (\vec{u} +
      \xi\vec v_i) \|_2^2)\Bigr|\\
      & \leq   \frac{1}{4}\sum_{\xi\in\{\pm 1,\pm i\}} \Bigl|
      \|\vec{u} + \xi\vec v_i \|_2^2  - \|\vec B (\vec{u} +
      \xi\vec v_i) \|_2^2\Bigr|\\
      & \leq \max_{\xi\in\{\pm 1,\pm i\}} \Bigl| 
      \|\vec{u} + \xi\vec v_i \|_2^2 - \|\vec B(\vec{u} +
      \xi\vec v_i) \|_2^2 \Bigr|
      =: m_i.
      \label{eq:ploarID}
    \end{split}
\end{equation}
Thus, we have
\begin{equation}
  \begin{split}
    \text{(i)}
    &\leq\Prob{\lVert \vec m\rVert_p>\frac{\varepsilon'-\rho_0}{y_0}}
    +\Prob{\lVert \vec a\rVert_q + \lVert \vec b\rVert_q\geq y_0}.
  \end{split}
\end{equation}     
Next, we use $p=\infty$, $q=1$, $\lVert \vec a\rVert_1\leq\sqrt{n}$ and
$\lVert \vec b\rVert_1\leq\sqrt{n}\lVert \vec b\rVert_2$. 
By assumption $\vec \Phi$ is isotropically distributed, i.e., each
component of $\vec{m}$ has the same distribution.
Thus, $\lVert \vec
m\rVert_\infty$ is the maximum over $4n$ identically distributed 
random variables. Define $u_0=y_0/\sqrt{n}-1$. Using  the union bound
and the concentration inequality \eqref{eq:concentration} we have,
\begin{equation}
  \begin{split}
    \text{(i)}
    &\leq 
    4n \Prob{|m_{1}| 
      > \frac{\varepsilon' - \rho_0}{y_0} } + \Prob{\lVert \vec b\rVert_2 \geq
      \frac{y_0}{\sqrt{n}}-1}\\
    &=
    4n \Prob{|m_{1}| 
      > \frac{\varepsilon' - \rho_0}{\sqrt{n}(u_0+1)} } + \Prob{\lVert
      \vec b\rVert_2 \geq
      u_0}\\
    &\leq 
    4n \exp(- \gamma(\frac{\varepsilon' - \rho_0 }{\|\vec{u} + \vec
      v_1 \|_2^2 \sqrt{n}(1+u_0)})) + \Prob{\lVert \vec b\rVert_2 \geq
      u_0}\\ 
    &\leq 
    4n \exp(- \gamma(\frac{\varepsilon' - \rho_0 }{4\sqrt{n}(1+u_0)}))
    + \Prob{\lVert \vec b\rVert_2 \geq
      u_0}\\   
    &\leq 
    4n \exp(- \gamma(\varepsilon)) + \Prob{\lVert \vec b\rVert_2 \geq
      u_0}\\    
  \end{split}
\end{equation}
The last steps follow from $\|\vec{u} + \vec v_1 \|_2^2\leq
4$ and with $\varepsilon'=\varepsilon 4  \sqrt{n}(1+u_0) + \rho_0$. 
Since $ \Prob{\|\vec \Psi \vec \Phi \vec{u}\|_2 > u_0}
\leq \Prob{s_{\max}(\vec \Psi \vec \Phi )> u_0}$ the claim follows
from the last equation and \eqref{eq:CS:step1}.
\end{IEEEproof}

Now we are ready to prove Theorem
\ref{theo:RateEst}.
\begin{IEEEproof}[Proof of Theorem \ref{theo:RateEst}]
  Let $\set S \subseteq \set N$ be arbitrary but fixed. By the
  assumptions the rate gap bound in Lemma \ref{lem:lipschitz} can be
  rewritten as
  \[
  \Delta_i(\set S) \leq 2P \|\vec h_i \|_2^2 \sum_{l\in \set S} \left|
    |\langle \vec{\bar h}_i , \vec e_l \rangle|^2 - |\langle \vec \Psi
    (\vec \Phi \vec{\bar h}_i + \vec{\bar \mu}_i) , \vec e_l \rangle|^2
  \right|,
  \]
  where we defined $ \vec{\bar h}_i = \vec{h}_i/ \| \vec{h}_i \|_2$
  and $\vec{\bar \mu}_i = \vec{\mu}_i/\| \vec{h}_i \|_2$. If we fix
  $|\set S| = n$, Lemma \ref{lem:motiNOISE} yields
  \begin{multline}
    \Prob{\Delta_i(\set S) > 2P \|\vec h_i \|_2^2 (4
      \sqrt{n}(1+u_0)\varepsilon + \rho_0)}\leq 4n
    \exp(-\gamma(\varepsilon)) + \\ \Prob{s_{\max}(\vec \Psi \vec
      \Phi) > u_0} + \Prob{\|\vec \Psi \vec{\bar \mu}_i \|_2\left(\|\vec
      \Psi \vec{\bar \mu}_i\|_2 + 2\|\vec \Psi\vec \Phi\vec{\bar h}_i
      \|_2 \right)> \rho_0}, 
  \end{multline}
  for an arbitrary $i\in \set S$. Taking the union bound over all $i \in \set S$ yields, 
  \begin{multline}
    \Prob{\exists i \in \set S : \Delta_i(\set S) > 2P \|\vec h_i \|_2^2 (4
      \sqrt{n}(1+u_0)\varepsilon + \rho_0)}\leq 4n^2
    \exp(-\gamma(\varepsilon)) + \\ n \Prob{s_{\max}(\vec \Psi \vec
      \Phi) > u_0} + \sum_{i \in \set S}\Prob{\|\vec \Psi \vec{\bar \mu}_i \|_2(\|\vec
      \Psi \vec{\bar \mu}_i\|_2 + 2\|\vec \Psi\vec \Phi\vec{\bar h}_i
      \|_2) > \rho_0}, 
  \end{multline}
  Finally, applying the union bound over all $\binom{N}{n}$
  scheduling decisions $\set S \subseteq \set N$, with $|\set S|=n$,
  \begin{multline}
    \Prob{\exists \set S \subset \set N , |\set S| = n, \exists i \in \set S  :\Delta_i(\set S) > 2P \|\vec h_i \|_2^2 (4
      \sqrt{n}(1+u_0)\varepsilon + \rho_0)} \\ \leq \exp(\log(4n^2) + n \log(Ne/n) - \gamma(\varepsilon)) 
    +\exp(n \log(Ne/n)) \Prob{s_{\max}(\vec \Psi \vec \Phi) >
      u_0 } \\ + \exp(n \log(Ne/n) )  \max_{i\in \set N}\Prob{\|\vec \Psi \vec{\bar \mu}_i \|_2(\|\vec
      \Psi \vec{\bar \mu}_i\|_2 + 2\|\vec \Psi\vec \Phi\vec{\bar h}_i
      \|_2) > \rho_0},
 \end{multline}
where we used $\binom{N}{n} \leq  (Ne/n)^n$.


\end{IEEEproof}


\subsection{Proof of Corollary \ref{coro:linEst}}
\label{proof:linEst}

The following result will be useful in the proof. Let
$\vec a$ be a random vector with elements $a_i \sim \set C \set N
(0,1)$. Then, for all $t>0$,
\begin{equation}
\label{eq:conGauss}
  \Prob{\| \vec a  \|_2^2 - \Ex{\| \vec a  \|_2^2} > t}  \leq \exp(-t^2 / 2). 
\end{equation}
In fact, this is a special case of the concentration of measure
theorem for Lipschitz functions, see \cite[Theorem 8.40]{Foucart2013}.

\begin{IEEEproof} 
  For an arbitrary but fixed $\vec h_i$. Setting $e_0=0$, we have 
\[\Prob{ \|\vec \Psi \vec e \|_2 ( \|\vec
    \Psi \vec e \|_2 + 2\| \vec \Psi \vec \Phi \vec u \|_2) > e_0}
  =0.\] Since $\vec \Phi^+ \vec \Phi$ is a projector (i.e. Hermitian
  and idempotent) $s_{\max} (\vec \Phi^+\vec \Phi)=1$, and therefore
  we can set $u_0 = 1$ and obtain $ \Prob{ s_{\max} (\vec \Psi \vec
    \Phi) > 1 } = 0.  $ Using \eqref{eq:gaussPhi} we get from Theorem
  \ref{theo:RateEst}
  \[
  \Prob{\exists \set S \subset \set N , |\set S| \leq n \leq N/2 : \Delta_i >  16P\|\vec h_i \|^2_2 \sqrt{n} \varepsilon')}
  \leq 4nN\left( \multi{N}{n} \right)\exp\left(-M \varepsilon'^2 / \kappa  \right),
  \]
  with $\kappa = \frac{2}{1 - \ln(2)}$.
  Since $\vec h_i$ is also random we can use Lemma \ref{lem:prob} and get
  \[
  \Prob{\exists i \in \set N : \Delta_i >  16P h_0 \sqrt{n} \varepsilon')}
  \leq  4nN\left( \multi{N}{n} \right)\exp\left(-M \varepsilon'^2 / \kappa  \right) +
  \Prob{\|\vec h_i \|^2_2 > h_0}.
  \]
  By assumption we have $\Ex{\|\vec h_i \|^2_2}=k$. Thus,
  \eqref{eq:conGauss} gives,
  \[
  \Prob{\|\vec h_i \|^2_2 > h_0} = \Prob{\|\vec h_i \|^2_2 > t+k} \leq \exp(-t^2/2).
  \]
  Hence, if we set $h_0 = t+k$ and $t = \sqrt{2M \varepsilon'^2 /
    \kappa}$, 
  \[
  \Prob{\exists i \in \set N : \Delta_i >  16P \sqrt{n} (\sqrt{2M \varepsilon'^2 /
    \kappa}+k) \varepsilon')}
  \leq \left(4nN\left( \multi{N}{n} \right)+1\right)\exp\left(-M \varepsilon'^2 / \kappa  \right).
  \]
  Finally, setting $\varepsilon = (4nN+1)\exp\left(-M \varepsilon'^2 /
    \kappa  \right)$ the claim follows. 
\end{IEEEproof}

\bibliographystyle{IEEEtran}
\bibliography{refs,references}

\end{document}